\documentclass[prd,byrevtex
,preprint
,tightenlines
,showkeys
]{revtex4}
\usepackage{amsmath,amssymb,amsxtra}
\usepackage{graphicx}
\newcommand{\del}{\delta}

\newcommand{\bk}{\boldsymbol{k}}
\newcommand{\bx}{\boldsymbol{x}}
\newcommand{\VEV}[1]{\langle #1 \rangle}
\newcommand{\0}[1]{\overset{(0)}{#1}}
\newcommand{\1}[1]{\overset{(1)}{#1}}
\newcommand{\2}[1]{\overset{(2)}{#1}}
\begin{document}
\title{The back reaction and the effective Einstein's equation for the Universe with ideal fluid cosmological perturbations}
\author{Yasusada Nambu}
\affiliation{Department of Physics, Graduate School of Science, Nagoya 
University, Chikusa, Nagoya 464-8602, Japan}
\email{nambu@allegro.phys.nagoya-u.ac.jp}
\date{February 1, 2002}
\begin{abstract}
    We investigate the back reaction of cosmological perturbations on 
    the evolution of the Universe using the renormalization group 
    method. Starting from the second order perturbed Einstein's 
    equation, we renormalize a scale factor of the Universe and 
    derive the evolution equation for the effective scale factor which includes back reaction due to inhomogeneities of the Universe. The resulting equation has the same form as the standard Friedman-Robertson-Walker equation with the effective energy density and pressure which represent the back reaction effect.
\end{abstract}
\maketitle

\section{introduction}

Owing to the nonlinear nature of Einstein's equation, fluctuations of
the metric  affect the evolution of the background
space time. In cosmology, this effect is expected to be important when we
consider the evolution of large scale nonlinear structures and the
evolution of the early Universe. This is cosmological the back
reaction problem and has been studied by several
authors\cite{issacson68,futamase89,futamase96,russ97,boersma98,
mukhanov97,abramo97a,abramo97b,abramo99,nambu00,nambu01}.
 
One of the main difficulty in the cosmological back reaction problem is 
connected with the gauge freedom of  perturbation variables. 
The back reaction effect
appears from the second order quantities in a perturbation expansion 
and we must use the second order gauge invariant quantities to avoid
the gauge ambiguity of the back reaction problem. Abramo and
co-workers\cite{mukhanov97,abramo97a,abramo97b,abramo99} derived the
gauge-invariant effective energy momentum tensor of cosmological
perturbations and applied their formalism to the inflationary
universe. They discussed the effect of the inhomogeneity on the
background Friedman-Robertson-Walker (FRW) universe, but did not derive solutions of an effective
scale factor for the FRW universe with the back reaction.

In our previous
papers\cite{nambu00,nambu01}, the renormalization group
method\cite{chen96,kunihiro95,nozaki99,nambu99} was applied to the 
cosmological back reaction problem. 
We start from the following perturbation expansion of the metric
\begin{equation}
  g_{ab}=\0{g}_{ab}+\1{g}_{ab}+\2{g}_{ab}+\cdots,
\end{equation}
where $\0{g}_{ab}$ is the background FRW metric and represents a homogeneous
and isotropic space. $\1{g}_{ab}$ is the metric of the first order
linear perturbation with $\VEV{\1{g}_{ab}}=0$ where $\VEV{\cdots}$ means
the spatial average with respect to the background FRW metric.
 $\2{g}_{ab}$ is the second order metric and this part 
contains nonlinear effects caused by the first order linear
perturbation. This nonlinearity produces homogeneous and isotropic
zero modes as part of the second order metric. That is,
\begin{equation}
  \VEV{\2{g}_{ab}}\neq 0.
\end{equation}
As the zero mode part of the metric has the same symmetry as the 
background FRW universe, it must be interpreted as a part of FRW metric.
Hence we  redefine the background metric as
follows:
\begin{equation}
  \overset{(B)}{g}_{ab}=\0{g}_{ab}+\VEV{\2{g}_{ab}}.
\end{equation}
We used the renormalization group method to define the new background
metric $\overset{(B)}{g}_{ab}$ which includes the back reaction
effect. In previous papers\cite{nambu00,nambu01}, we first solved
perturbation equations and obtained $\1{g}_{ab}$ and $\2{g}_{ab}$. 
Then the renormalized effective scale factor with the back reaction 
effect was obtained. 

But it is more convenient to obtain the renormalized Einstein's equation
for the zero mode variables from the first to investigate the back reaction
effect. This is equivalent to average Einstein's equation and derive
the evolution equation for the effective FRW universe with the back 
reaction effect. In this paper,
we apply the renormalization group method directly to the perturbed
Einstein's equation. We does not use an explicit form of the first and 
the second order solutions of perturbations. All that we need is how 
 constants of integration enter in the solution of perturbations.  

The plan of this paper is as follows. In Sec. II, we introduce the
gauge ready formalism of  cosmological perturbations and derive the
second order Einstein's equation for  gauge invariant
variables. In Sec. III, the renormalization group method is applied to
the zero mode part of the Einstein's equation and the evolution equation for
the renormalized scale factor is derived. In Sec. IV, we investigate the solution of the
back reaction equation. Sec. V is devoted for summary and
discussion. We use units in which $c=\hbar=8\pi G=1$ throughout the
paper.
 
\section{the second order perturbation for the Universe with ideal fluid}

To circumvent the gauge ambiguity of  cosmological perturbations 
and to obtain a gauge independent interpretation of the cosmological back 
reaction problem, we must use gauge invariant description of 
cosmological perturbations\cite{kodama84,mukhanov92}. Abramo and 
co-workers obtained the gauge invariant effective energy-momentum tensor for 
cosmological perturbations\cite{mukhanov97,abramo97a,abramo97b,abramo99}
 which is quadratic with respect to the first order variables. 
To apply the renormalization group method to the back reaction 
problem, we must obtain the second order zero mode perturbation in a 
gauge invariant manner\cite{nambu00,nambu01}. In this section, we 
introduce the gauge ready formalism\cite{hwang91} and  derive the 
evolution equation for the second 
order zero mode perturbation which is invariant under the first order 
gauge transformation.
\subsection{the gauge ready form of cosmological perturbations}
We consider a spatially flat FRW universe with perfect fluid of which 
equation of state is given by $p=w\,\rho$, where $w$ is assumed to be 
constant. The background scale factor of the Universe and the energy 
density are given by
\begin{equation}
    a(t)=C\,t^{\frac{2}{3(1+w)}},\quad 
    \0{\rho}=\frac{c_{0}}{a^{3(1+w)}},
    \label{eq:back}
\end{equation}
where $C$ and $c_{0}$ are constants of integration. The energy density of the fluid 
$\0{\rho}$ is obtained from the zeroth order of the following mass 
conservation equation
\begin{equation}
 \left(\rho^{\frac{1}{1+w}} \sqrt{-g}\,u^a\right)_{,a}=0, \label{eq:mass}
\end{equation}
where $u^{a}$ is four velocity of the fluid. 
We restrict our attention  only to the scalar 
type perturbation and the metric of the first order perturbation can 
be written as
\begin{equation}
    \1{g}_{ab}=
    \left(
    \begin{array}{ll}
        -2\phi & a(t) B_{,i} \\
        a(t) B_{,i} & a^{2}(t)\left[(1-2\psi)\,\del_{ij}+2E_{,ij}\right]
    \end{array}
    \right).
    \label{eq:metric1st}
\end{equation}
Let us consider the infinitesimal coordinate transformation
\begin{equation}
    x'{}^{0}=x^{0}+\xi^{0},\quad 
    x'{}^{i}=x^{i}+\frac{1}{a^{2}}\del^{ij}\xi_{,j}
\end{equation}
Under this transformation, the perturbation variables receive the 
following gauge transformations:
\begin{align}
    & \phi'=\phi-\dot\xi{}^{0},& & 
    B'=B-a\left(\frac{\xi}{a^2}\right)\spdot+\frac{\xi^0}{a}, \notag\\
    & \psi'=\psi+H \xi^0,& &E'=E-\frac{\xi}{a^2}.
\end{align}
The comoving three velocity $v_{i}=v_{,i}$ of the fluid transforms as
\begin{equation}
    v'=v+a^2 \left(\frac{\xi}{a^2}\right)\spdot,
    \label{eq:}
\end{equation}
and the velocity potential defined by $\chi\equiv v+a B$ transforms as
\begin{equation}
 \chi'=\chi+\xi^0.
\end{equation}
From these transformations, we can make the following gauge invariant 
combination of  perturbation variables:
\begin{align}
    &\Phi\equiv\phi+\left[a(B-a\dot E)\right]\spdot,& \quad &
    \Psi\equiv\psi-aH(B-a\dot E), \notag \\
 &X\equiv\chi-a(B-a\dot E),& \quad &
 \mathcal{R}\equiv\psi-H\chi.
\end{align}
Now the vector 
\begin{equation}
     \xi^\mu_{\text{C}}\equiv \left(-\chi,~ \del^{ij} E_{,j}\right)
    \label{eq:vecc}
\end{equation}
transforms as four vector under the coordinate transformation. The 
gauge transformation induced by the coordinate transformation using the 
vector \eqref{eq:vecc} is
\begin{align}
 &\phi'=\phi+\dot\chi=\Phi+\dot X,& &B'=B-a\dot 
 E-\frac{\chi}{a}=-\frac{X}{a}, \notag\\
 &E'=E-E=0,& &\psi'=\psi-H\chi=\mathcal{R}, \label{eq:comov}\\
 &\chi'=\chi-\chi=0.&& \notag
\end{align}
Hence in this gauge, each components of the metric correspond to the gauge 
invariant variables and a comoving gauge condition $\chi=0$ is satisfied. 
The choice of the vector \eqref{eq:vecc} is not unique. The vector
\begin{equation}
    \xi^{\mu}_{\text{L}}=\left(-a(B-a\dot E),~ \del^{ij}E_{,j} \right)
    \label{eq:vecl}
\end{equation}
also transforms as a four vector and the gauge transformation using 
this vector is
\begin{align}
 & \phi'=\phi+\left[a(B-a\dot E)\right]\spdot=\Phi,&
 &B'=0,& & E'=0,\\
 & \psi'=\psi-aH(B-a\dot E)=\Psi,& & \chi'=\chi-a(B-a\dot 
 E)=X.& &
\end{align}
This gauge corresponds to the longitudinal gauge and components of the metric 
coincide with  gauge invariant variables $\Phi$ and $\Psi$. This gauge 
is often used to investigate cosmological perturbations\cite{mukhanov92}, and Abramo and 
co-worker\cite{mukhanov97,abramo97a,abramo97b,abramo99} also used  this 
gauge to evaluate the gauge invariant energy momentum tensor of cosmological perturbations. 
But for the purpose of the back reaction problem, this gauge is not 
suitable because the contribution of long wavelength limit mode of the
perturbations to  the effective energy momentum tensor 
does not vanish automatically\cite{nambu01}. 
The perturbation of which wavelength is infinite cannot be 
distinguished from the background and has nothing to do with the back 
reaction effect. But in the longitudinal gauge, we have contribution of the 
long wavelength limit perturbation to the second order quantities. We must 
subtract this to obtain correct back reaction effect. 
On the other hand, in comoving gauge, the second order perturbation 
becomes zero in the long wavelength limit. 
In this paper, therefore, we adopt the 
gauge ready form in the comoving gauge to investigate the back 
reaction problem.

To simplify the calculation, we choose $B=0$ and the metric up to 
the second order in the  gauge \eqref{eq:comov} is given by
\begin{equation}
 ds^2=-(1+2\phi+2\phi_2)\,dt^2-2a^2\dot E_{,i}\,dx^i dt
 +a^2(t)(1-2\psi-2\psi_2)\,d\bx^2. \label{eq:ready}
\end{equation}
As the each component of the first order metric perturbation corresponds to the gauge invariant 
variable in this gauge, the second order zero mode variables 
$\phi_{2}$ and  $\psi_{2}$ determined by the first order quantities 
are also invariant under the first order 
gauge transformation up to the second order.

Using the metric \eqref{eq:ready} with the comoving condition 
$\chi=0$, the first order Einstein equations are
\begin{subequations}
\begin{align}
  & \frac{1}{a^2}\nabla^2\psi+H\nabla^2\dot E=\frac{1}{2}\,\1{\rho}, \\
  & \dot\psi+H\phi=0, \\
  & 
  \ddot\psi+3H\dot\psi+H\dot\phi-3wH^2\phi=\frac{w}{2}\,\1{\rho}, \\
  & \ddot E+3H\dot E+\frac{1}{a^2}(\psi-\phi)=0.
\end{align}
\end{subequations}
%
%
%
We can obtain the following evolution equation for the spatial curvature perturbation $\psi$
\begin{equation}
    \ddot\psi+3H\dot\psi-\frac{w}{a^2}\nabla^2\psi=0.
    \label{eq:}
\end{equation}
%

\subsection{the evolution equations for the second order zero mode perturbation}

The second order Einstein equations for the zero mode variables 
$\phi_{2}$ and $\psi_{2}$ are
\begin{subequations}
\begin{align}
 & 6H(\dot\psi_2+H\phi_2)=-\VEV{\2{G}{}^0{}_0(\1{g})}
   -\2{\rho}, \label{eq:2nd-a}\\
 &2[\ddot\psi_2+H(3\dot\psi_2+\dot\phi_2)-3H^2w\phi_2]
 =-\frac{1}{3}\VEV{\2{G}{}^{i}{}_i(\1{g})}+w\,\2{\rho}, 
 \label{eq:2nd-b}
\end{align}
\end{subequations}
where
\begin{multline}
\VEV{\2{G}{}^0{}_0(\1{g})}
=\Biggl\langle-3\dot\psi^2+12H\psi\dot\psi
-\frac{5}{a^2}\psi\nabla^2\psi-6H\psi\nabla^2\dot E \\
-3H^2a^2\dot E\nabla^2\dot E 
-2H\phi\left(-\nabla^2 \dot E
+6\dot\psi\right)-12H^2\phi^2\Biggr\rangle, 
\end{multline}
\begin{multline}
\VEV{\2{G}{}^{i}{}_j(\1{g})} 
=\Biggl\langle-\frac{5}{a^2}\psi\nabla^2\psi+\dot\psi^2
+4\psi(\ddot\psi+3H\dot\psi)-2\nabla^2\dot
E\left(3H\psi+\frac{2}{3}\dot\psi\right) \\
-2\psi\nabla^2\ddot E 
-2a^2H\dot E\nabla^2\ddot E
-(2\dot H+5H^2)a^2\dot E\nabla^2\dot E 
+\frac{2}{3}\dot E\nabla^2\dot\phi-2\dot\phi\dot\psi \\
 -4\phi\left(\ddot\psi+3H\dot\psi
-\frac{1}{2a^2}\nabla^2\psi-\frac{1}{3}\left(\nabla^2\ddot E
+2H\nabla^2\dot E\right)+2H\dot\phi\right) \\
-\frac{2}{3a^2}\phi\nabla^2\phi-4\phi^2\left(3H^2
+2\dot H\right)\Biggr\rangle\,\del{}^i{}_j.
\end{multline}
Eq.\eqref{eq:2nd-a} is a time-time component and Eq.\eqref{eq:2nd-b} 
is a space-space component of the spatially averaged Einstein's 
equation $\VEV{G^{a}{}_{b}}=\VEV{T^{a}{}_{b}}$. 
The spatial average is defined by
$$
  \VEV{A}=\frac{1}{V}\int d^3 x\,A,
$$
where $V$ is the volume of a sufficiently large compact domain and 
periodic boundary conditions for perturbations are assumed. The 
second order energy density $\2{\rho}$ 
is determined by the mass conservation \eqref{eq:mass} and given by
\begin{equation}
   \2{\rho}=(1+w)\,\0{\rho}\,(3\psi_2-\phi_2)+3\VEV{\psi\,\1{\rho}}
   +\frac{w}{2(1+w)}\frac{\VEV{\1{\rho}{}^2}}{\0{\rho}}+c_2(1+w)\,\0{\rho},
\end{equation}
where $c_{2}$ is a constant.

\section{the renormalization and the effective Einstein's  equation}
In this section, we derive the effective Einstein's equation which describes the evolution of 
the effective FRW universe with the back reaction effect using the renormalization group method. 
The spatially averaged Einstein's equation up to the second order is
\begin{subequations}
\begin{align}
 &-3H^2+\0{\rho}+6H(\dot\psi_2+H\phi_2)+(1+w)\,\0{\rho}(3\psi_2-\phi_2+c_2)
 =-\rho_{\text{BR}}, \label{eq:02a}\\
 &-2\dot H-3H^2-w\,\0{\rho}+2\left[\ddot\psi_2
 +H(3\dot\psi_2+\dot\phi_2)-3H^2w\phi_2\right] \notag \\
 &\qquad\qquad\qquad\qquad\qquad\qquad
 -w(1+w)\0{\rho}(3\psi_2-\phi_2+c_2)=p_{\text{BR}}, \label{eq:02b}
\end{align}
\end{subequations}
where
\begin{subequations}
\begin{align}
 &\rho_{\text{BR}}\equiv \VEV{\2{G}{}^0{}_0(\1{g})}+3\VEV{\psi\1{\rho}}
 +\frac{w}{2(1+w)}\frac{\VEV{\1{\rho}{}^2}}{\0{\rho}}, \\
 &p_{\text{BR}}\equiv -\frac{1}{3}\,\VEV{\2{G}{}^i{}_i(\1{g})}
 +3w\VEV{\psi\1{\rho}}+\frac{w^2}{2(1+w)}\frac{\VEV{\1{\rho}{}^2}}{\0{\rho}}.
\end{align}
\end{subequations}
By introducing a new time variable $d\tau\equiv (1+\phi_{2})\,dt$, 
the Einstein equations \eqref{eq:02a} and \eqref{eq:02b} become
\begin{subequations}
\begin{align}
&-3H^2+6H\dot\psi_2+\0{\rho}\left[1+(1+w)(3\psi_2-\phi_2+c_2)\right]
=-\rho_{\text{BR}}, \label{eq:naive-a}\\
 &-2\dot H-3H^2+2\,(\ddot\psi_2+3H\dot\psi_2)
 -w\0{\rho}\left[1+(1+w)(3\psi_2-\phi_2+c_2)\right]=p_{\text{BR}}, \label{eq:naive-b}
\end{align}
\end{subequations}
where $\dot{}=\frac{d}{d\tau}$.

The spatially averaged line element up to the second order perturbation is
\begin{equation}
 \VEV{ds^2}=-d\tau^2+a^2(\tau)\left[1-2\,\psi_2(\tau)
 +2\psi_2(\tau_0)\right]\,d\bx^2,\quad 
 a(\tau)=C\tau^{\frac{2}{3(1+w)}}, \label{eq:zerometric}
\end{equation}
where $\tau_{0}$ is a time at which the initial condition is set, and 
it is always possible to write the zero mode metric as 
Eq.\eqref{eq:zerometric} by choosing a constant 
of integration $c_{2}$ contained in $\2{\rho}$  as $c_{2}=-3\,\psi_{2}(\tau_{0})$. 
From this form of the line element, we can observe that the effective scale factor for the FRW universe is
\begin{equation}
  a_{\text{eff}}(\tau)=\tau^{\frac{2}{3(1+w)}}\,C\,\left[1-\psi_2(\tau)+\psi_2(\tau_0)\right].
  \label{eq:aeff}
\end{equation}
We regard the second order perturbation $\psi_{2}$ as a secular term and apply the
renormalization group method\cite{chen96,kunihiro95,nozaki99,nambu99} 
to absorb it to the zeroth order constant of integration $C$.
We redefine the zeroth order integration constant $C$ as
$$
 C=C(\mu)+\del C(\mu;\tau_0),
$$
where $\mu$ is a renormalization point and $\del C$ is a counterterm 
which absorbs the secular divergence of the solution. We assume that $\del C$ is the second order quantity. The effective 
scale factor \eqref{eq:aeff} up to the second order of the perturbation can be written
\begin{align*}
 a_{\text{eff}}(\tau)
 &=\tau^{\frac{2}{3(1+w)}}\,
 (C(\mu)+\del C(\mu;\tau_0))(1-\psi_2(\tau)+\psi_2(\mu)-\psi_2(\mu)+\psi_2(\tau_0)) \\
 &=\tau^{\frac{2}{3(1+w)}}\,C(\mu)(1-\psi_2(\tau)+\psi_2(\mu)),
\end{align*}
where we have chosen the counterterm $\del C$ so as to absorb the 
$(\psi_{2}(\mu)-\psi_{2}(\tau_{0}))$-dependent term:
\begin{equation}
 \del C(\mu;\tau_0)+C(\mu)(-\psi_2(\mu)+\psi_2(\tau_0))=0.
\end{equation}
This defines the renormalization transformation
\begin{equation}
 \mathcal{R}_{\mu-\tau_0} : C(\tau_0)\longmapsto C(\mu)
 =C(\tau_0)-C(\mu)\,\left[\psi_2(\mu)-\psi_2(\tau_0)\right],
\end{equation}
and this transformation forms Lie group. The renormalization group 
equation is obtained by differentiating the both side of the 
transformation with respect to $\mu$ and setting $\tau_{0}=\mu$:
\begin{equation}
 \frac{dC(\mu)}{d\mu}=-C(\mu)\frac{d\psi_2(\mu)}{d\mu},
\end{equation}
and its solution is
\begin{equation}
 C(\mu)=C(\tau_{0})\, e^{-\psi_2(\mu)+\psi_{2}(\tau_{0})}.
\end{equation}
By equating $\mu=\tau$ in the effective scale factor, the renormalized 
scale factor is given by
\begin{equation} 
  a_R(\tau)=\tau^{\frac{2}{3(1+w)}}\,C(\tau)
 =a(\tau)\,e^{-\psi_2(\tau)+\psi_{2}(\tau_{0})}.
\end{equation}
By substituting the relation 
$a(\tau)=a_{R}(\tau)e^{\psi_{2}(\tau)-\psi_{2}(\tau_{0})}$ 
into \eqref{eq:naive-a} and \eqref{eq:naive-b}, and keeping terms up to 
the second order of the perturbation, we obtain the following equations 
for the renormalized scale factor $a_{R}$:
\begin{subequations}
\begin{align}
 & -3H^2+\rho_0=-\rho_{\text{BR}}, \label{eq:bra}\\
 & -2\dot H-3H^2-w\rho_0=p_{\text{BR}}, \label{eq:brb}
\end{align}
\end{subequations}
where $H=\frac{\dot a_{R}}{a_{R}}$ and
$$
 \rho_0\equiv  
 \left[1-(1+w)\phi_{2}\right]\frac{c_{0}}{(a_{R})^{{3(1+w)}}}
 =\frac{c_{0}}{\left[(1+\phi_{2})\,a_{R}^{3}\right]^{1+w}}.
$$
\eqref{eq:bra} and \eqref{eq:brb} are the main result of this paper. 
The second order curvature variable $\psi_{2}$ disappears in the 
evolution equation by the procedure of the renormalization. 
The second order lapse function $\phi_{2}$ remains in the expression $\rho_{0}$, but this variable 
corresponds to the second order gauge freedom to parameterize time and 
we can freely choose the form of this function. Eqs.\eqref{eq:bra} 
and \eqref{eq:brb} have the same form as the standard FRW equations. 
On the right-hand side, the back reaction effect appears as source 
terms $\rho_{\text{BR}}$ and $p_{\text{BR}}$, and their explicit form is determined by solving the first order perturbation.

\section{solutions of the effective Einstein's equation}

In this section, we solve the effective Einstein equations \eqref{eq:bra}
and \eqref{eq:brb}, and investigate how the inhomogeneity affects the 
expansion of the Universe. By eliminating the variable $\rho_{0}$, we have
\begin{equation}
 -2\dot H-3(1+w)H^2=w\rho_{\text{BR}}+p_{\text{BR}}.
\end{equation}
For $w\ne -1$, by using a new variable $x\equiv 
\left(a_{R}\right){}^{\frac{3}{2}(1+w)}$, 
\begin{equation}
 \ddot x=-\frac{3}{4}(1+w)(w\rho_{\text{BR}}+p_{\text{BR}})\,x. \label{eq:solve}
\end{equation}
We solve this equation for various value of $w$.
\subsection {vacuum energy case $w=-1$}
The universe expands exponentially $a\propto e^{Ht}, H=\text{const.}$, and we must treat this case separately. In this case, the perturbation of energy density is zero and the effective Einstein's equation becomes
\begin{align}
 & -3H^2+\rho_0=-\rho_{\text{BR}}=-\VEV{\overset{(2)}{G}{}^0{}_0(\overset{(1)}{g})}, \notag \\
 & -2\dot H-3H^2+\rho_0=p_{\text{BR}}=-\frac{1}{3}\VEV{\overset{(2)}{G}{}^i{}_i(\overset{(1)}{g})},\quad \rho_0=\text{const.}
\end{align}

For long wavelength perturbation $k\ll aH$, the first order growing mode solution is given by
$$
 \psi\approx\sum_{\bk}C_{\bk}e^{i\,\bk\cdot\bx},\quad \phi\approx0,\quad\dot E\approx-\frac{1}{Ha^2}\sum_{\bk}C_{\bk}e^{i\,\bk\cdot\bx},
$$
where $C_{\bk}$ is a constant in time. 
The source terms of the effective Einstein's equation are
$$
 \rho_{\text{BR}}\approx\frac{2}{a_R^2}\sum_{\bk}k^2|C_{\bk}|^2,\quad p_{\text{BR}}\approx-\frac{2}{3a_R^2}\sum_{\bk}k^2|C_{\bk}|^2=-\frac{1}{3}\rho_{\text{BR}}.
$$
This is equivalent to a negative spatial curvature term. For the slow roll inflationary phase driven by a scalar field, it was shown that the back reaction effect by the long wavelength perturbation is equivalent to  positive spatial curvature\cite{nambu01}. For vacuum energy case, fluctuation of the energy density becomes identically zero and this leads to the different back reaction effect compared to the inflation with a scalar field.

For short wavelength mode $k\gg aH$, the first order curvature perturbation is given by
$$
 \psi\approx\sum_{\bk}\frac{C_{\bk}}{ka}\cosh\left(\frac{k}{aH}\right)e^{i\bk\cdot\bx}
$$
and
$$
\rho_{\text{BR}}\approx\frac{7}{2}\sum_{\bk}\frac{|C_{\bk}|^2}{a_R^4},\quad p_{\text{BR}}\approx \frac{7}{6}\sum_{\bk}\frac{|C_{\bk}|^2}{a_R^4}.
$$
The back reaction effect is equivalent to a radiation fluid and this result is  the same as the inflation with a scalar field.

\subsection{dust case $w=0$}

The exact solution of the first order growing mode is 
\begin{equation}
 \psi=\sum_{\bk}C_{\bk}\,e^{i\,\bk\cdot\bx},\quad \phi=0,
 \quad \dot E=-\frac{3}{5}\tau^{-1/3}\sum_{\bk}C_{\bk}\,e^{i\,\bk\cdot\bx},
 \quad \1{\rho}=-\frac{6}{5}\sum_{\bk}\frac{k^2}{a{}^2}\,C_{\bk}\,e^{i\,\bk\cdot\bx},
\end{equation}
where $C_{\bk}$ is a constant in time. Using this solution, the source terms of the effective Einstein equation are
\begin{equation}
 \rho_{\text{BR}}=-\frac{13}{25}\sum_{\bk}\frac{k^2\,|C_{\bk}|^2}{a_{R}{}^2},\quad
  p_{\text{BR}}=\frac{13}{3\cdot25}\sum_{\bk}\frac{k^2\,|C_{\bk}|^2}{a_{R}{}^2}
 =-\frac{1}{3}\rho_{\text{BR}}.
\end{equation}
This is the same equation of state as a positive spatial curvature term. 
The  equation \eqref{eq:solve} becomes
\begin{equation}
 \ddot x=-\frac{3A}{4}\,x^{-1/3},\quad A
 =\frac{13}{3\cdot 25}\sum_{\bk}k^2|C_{\bk}|^2,
\end{equation}
and the solution is
\begin{equation}
  a_R=\frac{4E}{9A}(1-\cos\eta), \quad
  t=\frac{4E}{9A^{3/2}}(\eta-\sin\eta),
\end{equation}
where $E$ is a constant of integration. This solution is the same as 
a closed FRW universe with dust. Therefore, the inhomogeneity works 
as a positive spatial curvature in the dust dominated 
universe\cite{russ97,nambu00} and the Universe will recollapse.

\subsection{$w\neq 0$ case}
For long wavelength perturbations $ k\ll a\,H$, 
using the long wavelength expansion, the first order solution up to 
$O(k^{2})$ is given by
\begin{align}
 &\psi\approx\sum_{\bk}C_{\bk}\,e^{i\,\bk\cdot\bx},\quad \phi\approx0
 ,\quad \dot E\approx-\frac{3(1+w)}{3w+5}\,\tau^{\frac{3w-1}{3(1+w)}}
 \,\sum_{\bk}C_{\bk}\,e^{i\,\bk\cdot\bx}, \notag \\
 & \1{\rho}
 \approx-\frac{6(1+w)}{3w+5}\sum_{\bk}\frac{k^2}{a{}^2}\,C_{\bk}\,e^{i\,\bk\cdot\bx},
\end{align}
and
\begin{align}
 &\rho_{\text{BR}}\approx-\frac{9w^2+30w+13}{(3w+5)^2}
 \,\sum_{\bk}\frac{k^2}{a_{R}{}^2}\,|C_{\bk}|^2, \notag\\
 &p_{\text{BR}}\approx -\frac{162w^3+423w^2+276w-13}{3(3w+5)^2}
 \,\sum_{\bk}\frac{k^2}{a_{R}{}^2}\,|C_{\bk}|^2.
\end{align}
The effective Einstein equation \eqref{eq:solve}  becomes
\begin{equation}
 \ddot x=\frac{3}{4}(1+w)(wA-B)\,x^{\frac{3w-1}{3(w+1)}},
\end{equation}
where
\begin{equation}
 A=-\frac{9w^2+30w+13}{(3w+5)^2}\sum_{\bk}k^2|C_{\bk}|^2,\quad 
 B=-\frac{162w^3+423w^2+240w-13}{3(3w+5)^2}\sum_{\bk}k^2|C_{\bk}|^2.
\end{equation}
By integrating the equation with respect to $\tau$, we obtain
\begin{equation}
 \frac{\dot x^2}{2}+V_{\text{eff}}(x)=E,\quad 
 V_{\text{eff}}=
 -\frac{3}{8}\frac{(w+1)^2(135w^3+333w^2+201w-13)}{(3w+1)(3w+5)^2}
 \sum_{\bk}k^2|C_{\bk}|^2\,x^{\frac{2(3w+1)}{3(w+1)}},
\end{equation}
where $E$ is a constant of integration. If there is no inhomogeneity, $V_{\text{eff}}=0$ and the solution of the effective Einstein's equation reduces to that of the background FRW solution.  The condition $E>0$ is 
necessary to reproduce the background the solution when the 
inhomogeneity vanishes. We can obtain qualitative behavior of the effective scale factor by observing the shape of the effective potential $V_{\text{eff}}(x)$ (FIG. 1). 
\begin{figure}[!]
\includegraphics[width=\linewidth]{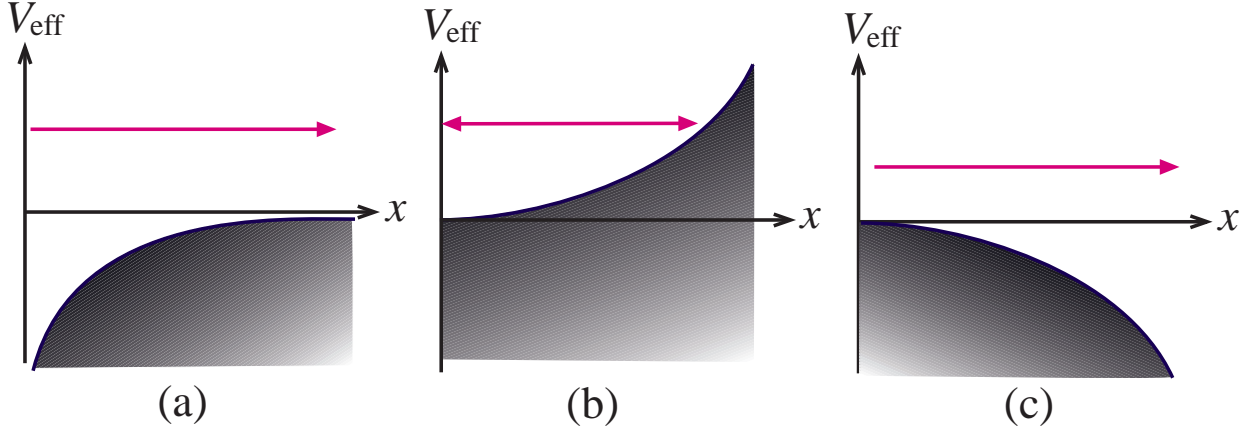}
\caption{Effective potentials for the long wavelength perturbations. (a) is for $-1<w<-1/3$, (b) is for $-1/3<w<w_*$ and (c) is for $w_*<w<1$. For $-1<w<w_*$, inhomogeneity reduces  the expansion rate of the Universe. For $w_*<w<1$, inhomogeneity enhances the expansion of the Universe. For $w=w_*$, there is no back reaction effect. }
\end{figure}
We can summarize the back reaction effect of the long wavelength perturbation on the expansion of the Universe as follows:
\begin{itemize}
 \item $-1<w<-1/3$ : the expansion of the Universe is accelerating. The 
 inhomogeneity reduces the expansion rate.
 \item $w=-1/3$ : $V_{\text{eff}}\propto \ln x$. The expansion of the Universe is  decelerating. The inhomogeneity reduces the expansion rate and the 
 universe will recollapse due to the back reaction effect.
 \item $-1/3<w<w_{*}$ : $w_{*}$ is a real root of the equation 
 $135w^3+333w^2+201w-13=0$ and $w_{*}\approx 0.06$. The expansion of the Universe is 
 decelerating. The inhomogeneity reduces the expansion rate and the 
 universe will recollapse.
 \item $w=w_{*}$ : $V_{\text{eff}}=0$ and there is no back reaction.
 \item $w_{*}<w<1$ : the expansion of the Universe is decelerating. 
 The inhomogeneity increases the expansion rate. 
 \end{itemize}
                  
For short wavelength perturbations $k\gg a\,H$, WKB type solution for the curvature perturbation $\psi$ is given by
\begin{equation}
 \psi\approx\begin{cases}
    \displaystyle\sum_{\bk}\frac{C_{\bk}}{k\sqrt{w}\,a}
    \cos\left(k\sqrt{w}\int\frac{d\tau}{a}\right)\,e^{i\,\bk\cdot\bx} \quad &\text{for}~~ w>0,\\
    \displaystyle\sum_{\bk}\frac{C_{\bk}}{k\sqrt{-w}\,a}
    \cosh\left(k\sqrt{-w}\int\frac{d\tau}{a}\right)\,e^{i\,\bk\cdot\bx} \quad &\text{for}~~ w<0,
   \end{cases}
\end{equation}
and using this solution, we have
\begin{equation}
 \rho_{\text{BR}}\approx 
 \mathrm{sign}(w)\frac{5-9w}{4w}\sum_{\bk}\frac{|C_{\bk}|^2}{a_{R}{}^4},
 \quad p_{\text{BR}}\approx \mathrm{sign}(w)\frac{5-9w}{12w}
 \sum_{\bk}\frac{|C_{\bk}|^2}{a_{R}{}^4}.
\end{equation}
The equation of state for the back reaction term becomes
\begin{equation}
 p_{\text{BR}}\approx \frac{1}{3}\rho_{\text{BR}}
\end{equation}
which is independent of  the value $w$. For $w<5/9$, $\rho_{\text{BR}}$ is positive and this is equivalent to a radiation fluid. The effective Einstein's equation becomes
\begin{equation}
 \frac{\dot x^2}{2}+V_{\text{eff}}(x)=E,\quad
 V_{\text{eff}}=-\mathrm{sign}(w)\frac{3}{32}\frac{(w+1)^2(5-9w)}{w}
 \sum_{\bk}|C_{\bk}|^2\,x^{\frac{2(3w-1)}{3(w+1)}},\quad E>0.
\end{equation}
%
\begin{figure}[!]
\includegraphics[width=\linewidth]{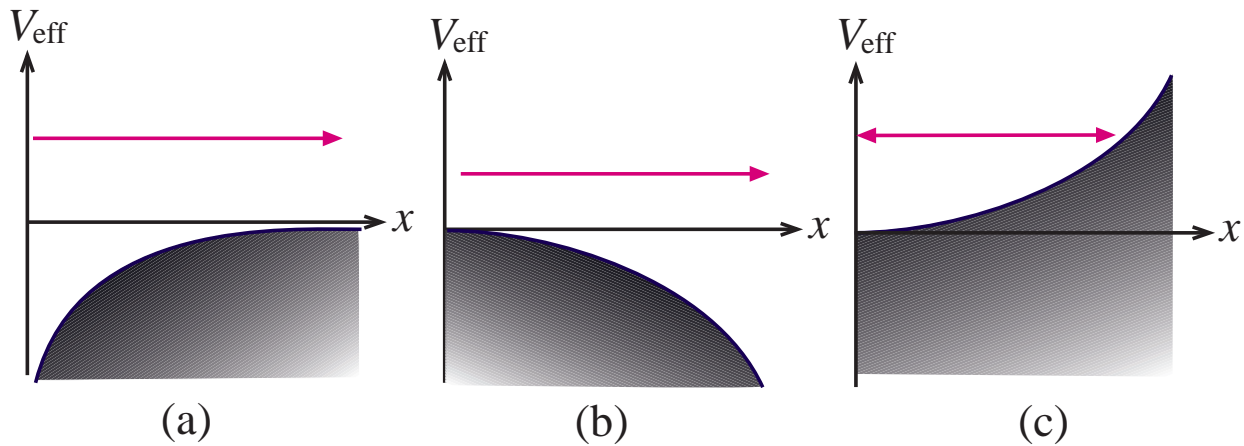}
\caption{Effective potentials for the short wavelength perturbations. (a) is for $-1<w<1/3$, (b) is for $1/3<w<5/9$ and (c) is for $5/9<w<1$. For $-1<w<1/3,5/9<w<1$, inhomogeneity reduces the expansion rate of the Universe. For $1/3<w<5/9$, inhomogeneity enhances the expansion of the Universe. For $w=1/3,5/9$, there is no back reaction effect.}
\end{figure}
The back reaction effect of the short wavelength perturbation on the expansion of the Universe is summarized as follows (FIG.~2):
\begin{itemize}
 \item $-1<w<1/3$ :  The inhomogeneity reduces the expansion rate.
 \item $w=1/3$ : $V_{\text{eff}}=0$. The equation of state of the back reaction term is the same as the back ground matter field and we have no  back reaction from inhomogeneity.
 \item $1/3<w<5/9$ : The inhomogeneity increases the expansion rate. 
 \item $w=5/9$ : $V_{\text{eff}}=0$. There is no back reaction.
 \item $5/9<w<1$ : The inhomogeneity reduces the expansion rate and the Universe will recollapse.
 \end{itemize}
%

\section{summary and discussion}
   In this paper,  we applied the renormalization group method to the perturbed Einstein's equation and  derived the evolution equation for the renormalized scale factor. The renormalized scale factor includes the back reaction effect caused by the inhomogeneity of the first order perturbation. The resulting equation has the same form as the standard FRW equation with the back reaction effect appears as the effective energy density and pressure on the right hand side of the equation. For the long wavelength perturbation, the equation of the state of the back reaction terms depends on the value $w$.  We have $p_{\text{BR}}=-\rho_{\text{BR}}/3, \rho_{\text{BR}}>0$ for $w=-1$ and this is equivalent to a negative spatial curvature. For $w=0$, we have $p_{\text{BR}}=-\rho_{\text{BR}}/3, \rho_{\text{BR}}<0$ and this is equivalent to a positive spatial curvature. On the other hand, for the short wavelength perturbation, the equation of state of the back reaction term becomes independent of $w$ and $p_{\text{BR}}\approx\rho_{\text{BR}}/3$. For $w<5/9$, $\rho_{\text{BR}}>0$ and the equation of state is the same as a radiation fluid.

   We derived the effective FRW equation \eqref{eq:bra} and \eqref{eq:brb} without using the explicit form of the background solution and the perturbation solution. Usually, the renormalization group method  is utilized to obtain the globally valid approximated solution of the  differential equation and we must prepare naive perturbative solution of the original equation before applying the procedure of renormalization. In this paper, we could obtain the evolution equation for the renormalized scale factor without solving the perturbation equations.  Once the effective FRW equation is obtained, it is possible to investigate the evolution of the effective scale factor and the back reaction effect by numerical method. Hence this approach is useful for the cosmological model with the scalar field in which case obtaining the solution of the perturbation is not easy in general. 

\vspace{1cm}

\noindent
{\bf ACKNOWLEDGMENT} \hfill

This work was supported in part by a Grant-In-Aid for Scientific
Research of the Ministry of Education, Science, Sports, and Culture of
Japan (11640270).


\begin{thebibliography}{18}
\expandafter\ifx\csname natexlab\endcsname\relax\def\natexlab#1{#1}\fi
\expandafter\ifx\csname bibnamefont\endcsname\relax
  \def\bibnamefont#1{#1}\fi
\expandafter\ifx\csname bibfnamefont\endcsname\relax
  \def\bibfnamefont#1{#1}\fi
\expandafter\ifx\csname citenamefont\endcsname\relax
  \def\citenamefont#1{#1}\fi
\expandafter\ifx\csname url\endcsname\relax
  \def\url#1{\texttt{#1}}\fi
\expandafter\ifx\csname urlprefix\endcsname\relax\def\urlprefix{URL }\fi
\providecommand{\bibinfo}[2]{#2}
\providecommand{\eprint}[2][]{\url{#2}}

\bibitem[{\citenamefont{Issacson}(1968)}]{issacson68}
\bibinfo{author}{\bibfnamefont{R.~A.} \bibnamefont{Issacson}},
  \bibinfo{journal}{Phys. Rev.} \textbf{\bibinfo{volume}{166}},
  \bibinfo{pages}{1263} (\bibinfo{year}{1968}).

\bibitem[{\citenamefont{Futamase}(1989)}]{futamase89}
\bibinfo{author}{\bibfnamefont{T.}~\bibnamefont{Futamase}},
  \bibinfo{journal}{Mon. Not. R. astr. Soc.} \textbf{\bibinfo{volume}{237}},
  \bibinfo{pages}{187} (\bibinfo{year}{1989}).

\bibitem[{\citenamefont{Futamase}(1996)}]{futamase96}
\bibinfo{author}{\bibfnamefont{T.}~\bibnamefont{Futamase}},
  \bibinfo{journal}{Phys. Rev. D} \textbf{\bibinfo{volume}{53}},
  \bibinfo{pages}{681} (\bibinfo{year}{1996}).

\bibitem[{\citenamefont{Russ et~al.}(1997)\citenamefont{Russ, Soffel, Kasai,
  and B{\"o}rner}}]{russ97}
\bibinfo{author}{\bibfnamefont{H.}~\bibnamefont{Russ}},
  \bibinfo{author}{\bibfnamefont{M.~H.} \bibnamefont{Soffel}},
  \bibinfo{author}{\bibfnamefont{M.}~\bibnamefont{Kasai}}, \bibnamefont{and}
  \bibinfo{author}{\bibfnamefont{G.}~\bibnamefont{B{\"o}rner}},
  \bibinfo{journal}{Phys. Rev. D} \textbf{\bibinfo{volume}{56}},
  \bibinfo{pages}{2044} (\bibinfo{year}{1997}).

\bibitem[{\citenamefont{Boersma}(1998)}]{boersma98}
\bibinfo{author}{\bibfnamefont{J.~P.} \bibnamefont{Boersma}},
  \bibinfo{journal}{Phys. Rev. D} \textbf{\bibinfo{volume}{57}},
  \bibinfo{pages}{798} (\bibinfo{year}{1998}).

\bibitem[{\citenamefont{Mukhanov et~al.}(1997)\citenamefont{Mukhanov, Abramo,
  and Brandenberger}}]{mukhanov97}
\bibinfo{author}{\bibfnamefont{V.~M.} \bibnamefont{Mukhanov}},
  \bibinfo{author}{\bibfnamefont{L.~R.} \bibnamefont{Abramo}},
  \bibnamefont{and} \bibinfo{author}{\bibfnamefont{R.~H.}
  \bibnamefont{Brandenberger}}, \bibinfo{journal}{Phys. Rev. Lett.}
  \textbf{\bibinfo{volume}{78}}, \bibinfo{pages}{1624} (\bibinfo{year}{1997}).

\bibitem[{\citenamefont{Abramo et~al.}(1997)\citenamefont{Abramo,
  Brandenberger, and Mukhanov}}]{abramo97a}
\bibinfo{author}{\bibfnamefont{L.~R.} \bibnamefont{Abramo}},
  \bibinfo{author}{\bibfnamefont{R.~H.} \bibnamefont{Brandenberger}},
  \bibnamefont{and} \bibinfo{author}{\bibfnamefont{V.~M.}
  \bibnamefont{Mukhanov}}, \bibinfo{journal}{Phys. Rev. D}
  \textbf{\bibinfo{volume}{56}}, \bibinfo{pages}{3248} (\bibinfo{year}{1997}).

\bibitem[{\citenamefont{Abramo}(1997)}]{abramo97b}
\bibinfo{author}{\bibfnamefont{L.~R.} \bibnamefont{Abramo}}, Ph.D. thesis,
  \bibinfo{school}{Brown University} (\bibinfo{year}{1997}),
  \bibinfo{note}{bROWN-HET-1096, gr-qc/9709049}.

\bibitem[{\citenamefont{Abramo}(1999)}]{abramo99}
\bibinfo{author}{\bibfnamefont{L.~R.} \bibnamefont{Abramo}},
  \bibinfo{journal}{Phys. Rev. D} \textbf{\bibinfo{volume}{D60}},
  \bibinfo{pages}{064004} (\bibinfo{year}{1999}).

\bibitem[{\citenamefont{Nambu}(2000)}]{nambu00}
\bibinfo{author}{\bibfnamefont{Y.}~\bibnamefont{Nambu}},
  \bibinfo{journal}{Phys. Rev. D} \textbf{\bibinfo{volume}{62}},
  \bibinfo{pages}{104010} (\bibinfo{year}{2000}),
  \bibinfo{note}{gr-qc/0006031}.

\bibitem[{\citenamefont{Nambu}(2001)}]{nambu01}
\bibinfo{author}{\bibfnamefont{Y.}~\bibnamefont{Nambu}},
  \bibinfo{journal}{Phys. Rev. D} \textbf{\bibinfo{volume}{63}},
  \bibinfo{pages}{044013} (\bibinfo{year}{2001}).

\bibitem[{\citenamefont{Chen et~al.}(1996)\citenamefont{Chen, Goldenfeld, and
  Oono}}]{chen96}
\bibinfo{author}{\bibfnamefont{L.-Y.} \bibnamefont{Chen}},
  \bibinfo{author}{\bibfnamefont{N.}~\bibnamefont{Goldenfeld}},
  \bibnamefont{and} \bibinfo{author}{\bibfnamefont{Y.}~\bibnamefont{Oono}},
  \bibinfo{journal}{Phys. Rev. E} \textbf{\bibinfo{volume}{54}},
  \bibinfo{pages}{376} (\bibinfo{year}{1996}).

\bibitem[{\citenamefont{Kunihiro}(1995)}]{kunihiro95}
\bibinfo{author}{\bibfnamefont{T.}~\bibnamefont{Kunihiro}},
  \bibinfo{journal}{Prog. Theor. Phys.} \textbf{\bibinfo{volume}{94}},
  \bibinfo{pages}{503} (\bibinfo{year}{1995}).

\bibitem[{\citenamefont{Goto et~al.}(1999)\citenamefont{Goto, Masutomi, and
  Nozaki}}]{nozaki99}
\bibinfo{author}{\bibfnamefont{S.}~\bibnamefont{Goto}},
  \bibinfo{author}{\bibfnamefont{Y.}~\bibnamefont{Masutomi}}, \bibnamefont{and}
  \bibinfo{author}{\bibfnamefont{K.}~\bibnamefont{Nozaki}},
  \bibinfo{journal}{Prog. Theor. Phys.} \textbf{\bibinfo{volume}{102}},
  \bibinfo{pages}{471} (\bibinfo{year}{1999}).

\bibitem[{\citenamefont{Nambu and Yamaguchi}(1999)}]{nambu99}
\bibinfo{author}{\bibfnamefont{Y.}~\bibnamefont{Nambu}} \bibnamefont{and}
  \bibinfo{author}{\bibfnamefont{Y.}~\bibnamefont{Yamaguchi}},
  \bibinfo{journal}{Phys. Rev. D} \textbf{\bibinfo{volume}{60}},
  \bibinfo{pages}{104011} (\bibinfo{year}{1999}).

\bibitem[{\citenamefont{Kodama and Sasaki}(1984)}]{kodama84}
\bibinfo{author}{\bibfnamefont{H.}~\bibnamefont{Kodama}} \bibnamefont{and}
  \bibinfo{author}{\bibfnamefont{M.}~\bibnamefont{Sasaki}},
  \bibinfo{journal}{Prog. Theor. Phys. Suppliment}
  \textbf{\bibinfo{volume}{78}}, \bibinfo{pages}{1} (\bibinfo{year}{1984}).

\bibitem[{\citenamefont{Mukhanov et~al.}(1992)\citenamefont{Mukhanov, Feldman,
  and Brandenberger}}]{mukhanov92}
\bibinfo{author}{\bibfnamefont{V.~F.} \bibnamefont{Mukhanov}},
  \bibinfo{author}{\bibfnamefont{H.~A.} \bibnamefont{Feldman}},
  \bibnamefont{and} \bibinfo{author}{\bibfnamefont{R.~H.}
  \bibnamefont{Brandenberger}}, \bibinfo{journal}{Phys. Rep.}
  \textbf{\bibinfo{volume}{215}}, \bibinfo{pages}{203} (\bibinfo{year}{1992}).

\bibitem[{\citenamefont{Hwang}(1991)}]{hwang91}
\bibinfo{author}{\bibfnamefont{J.}~\bibnamefont{Hwang}}, \bibinfo{journal}{ApJ}
  \textbf{\bibinfo{volume}{375}}, \bibinfo{pages}{443} (\bibinfo{year}{1991}).

\end{thebibliography}

\end{document}